\documentclass[a4paper,10pt]{article}

\usepackage[utf8x]{inputenc}

\usepackage{amsmath,amssymb,bm}
\usepackage{sidecap}
\usepackage{times}
\usepackage{graphicx}
\usepackage{float}
\usepackage[left=2.5cm,top=2.5cm,right=2.5cm,bottom=2.5cm,nohead]{geometry}

\usepackage{xcolor} 

\date{}              

\begin{document}

\begin{centering}
\LARGE\textbf{Controlled Coulomb effects in core-shell quantum rings}
       
\vspace{12pt}      
\normalsize\textbf{Anna Sitek$^{\ast,\ddagger}$, Miguel Urbaneja Torres$^{\ast}$, Kristinn Torfason$^{\ast}$,  Vidar Gudmundsson$^{\P}$, \\ and Andrei Manolescu$^{\ast}$}

\vspace{0pt}  
\normalsize\textit{$^{\ast}$School of Science and Engineering, Reykjavik University, Menntavegur 1, IS-101 Reykjavik, Iceland}\\
\normalsize\textit{$^{\ddagger}$Department of Theoretical Physics, Faculty of Fundamental Problems of Technology, Wroclaw University of Science and Technology, 50-370 Wroclaw, Poland}\\
\normalsize\textit{$^{\P}$Science Institute, University of Iceland, Dunhaga 3, IS-107 Reykjavik, Iceland}\\
\normalsize\textit{e-mail: annams@ru.is}\\
\end{centering}
\vspace{6pt}

\noindent
\textbf{ABSTRACT}\\
We analyse theoretically the possibilities of contactless control of in-gap states 
formed by a pair of electrons confined in a triangular quantum ring. The in-gap 
states are corner-localized states associated with two electrons occupying 
the same corner area, and thus shifted to much higher energies than other corner 
states, but still they are below the energies of corner-side-localized states.
We show how the energies, degeneracy and splittings between consecutive levels change 
with the orientation of an external electric field relatively to the polygonal 
cross section. We also show how absorption changes in the presence of 
external electric and magnetic fields.


\vspace{12pt}
\noindent
\textbf{1. INTRODUCTION}
\vspace{3pt}

\noindent
Core-shell semiconductor nanowires represent a novel expanding field of research 
largely as a result of their rich electronic properties, related both to transport 
and optics, and constitute a great technological promise as building blocks of 
electronic nanodevices. Such nanowires consist of an embedded core covered by 
different materials and are typically grown vertically by the bottom-up methods. 
Due to the crystallographic structure, the wires exhibit polygonal cross 
sections, most often hexagonal 
although other shapes, like circular, square 
or triangular, are also feasible. 
The possibility to combine vertically two different 
materials allows for control of band alignment, and thus structures in which 
electrons are accumulated only in the shells may be achieved. 
Moreover, the polygonal nanotubes of finite thickness
can be created by etching the core parts.

Core-shell nanowires short enough to confine the electrons along the growth 
direction, i.e. shorter than their wavelength, can be considered as quantum rings.
Such structures are of particular interest due to the 
possibility of controlling the number of carriers in the system and its 
fluctuations \cite{Ballester12, Ballester13}, and offer the possibility to study artificial 
atoms and their features as they exhibit shell structures similar to those of real 
atoms.

In this paper we focus on the manipulation possibilities of the in-gap states \cite{Sitek17} 
created by a pair of electrons confined in a triangular ring.
We show that an external electric field splits these states and 
moves them to different parts of the energy interval forbidden for non-interacting 
particles. It also considerably changes particle localization which depends on the 
field orientation relatively to the polygonal cross section.
We explain why in the presence of the electric field parallel to one of the sides 
two optical transitions to energetically separated in-gap states take place, 
and why they are blocked when the  
ring is immersed in external (static) magnetic field.

\vspace{12pt}
\noindent
\textbf{2. THE MODEL}
\vspace{3pt}

\noindent
We study a system of two Coulomb interacting electrons confined in a triangular
quantum ring. The sample is exposed to an external magnetic field $B$ perpendicular 
to the ring's plane ($x$,$y$), i.e., associated with a vector potential 
$\bm{A}=B[-y,x,0]/2$. The sample is also subjected to an external electric field 
which can be rotated with respect to the cross section.
We assume that it forms an angle $\phi$ with the $x$ axis, 
$\bm{E}=E[\cos(\phi),\sin(\phi),0]$.
The single-particle Hamiltonian is then 
\begin{eqnarray}
\label{Hamiltonian_single}
 H = \frac{\left(-i\hbar\nabla+e\bm{A}\right)^{2}}{2m_{\mathrm {eff}}}
 -g_{\mathrm {eff}}\mu_{\mathrm{B}}\sigma_{z}B-e\bm{E}\!\cdot\!\bm{r}\,,
\nonumber \\
\end{eqnarray}
where $m_{\mathrm {eff}}$ is the electron effective mass, $g_{\mathrm {eff}}$ 
is the effective g-factor,
$e$ the electron charge, $\sigma_{z}$ stands for the $z$ Pauli matrix and $\bm{r}$ 
defines the position.

To calculate single-particle eigenvalues ($E_{a}$) and eigenstates  
($\vert\psi_{a}\rangle$) we use a discretization method based on a polar
grid \cite{Daday11}. We construct a circular sample on which we superimpose 
polygonal constraints and redefine the grid such that it consists only of 
the sites situated 
between the boundaries. We solve the eigenvalue problem numerically and further 
use the results to construct the many-body Hamiltonian,
\begin{equation*}
\label{Hamiltonian_many}
\hat{H}=\sum\limits_{a}E_{a}a^{\dagger}_{a}a_{a}
+\frac{1}{2}\sum\limits_{abcd}V_{abcd}a^{\dagger}_{a}a^{\dagger}_{b}a_{d}a_{c}\,,
\end{equation*}
where $a^{\dagger}_{a}$ and $a_{a}$ create and annihilate, respectively, 
an electron in the state $\vert \psi_{a}\rangle$ and $V_{abcd}$ are the Coulomb 
integrals.

We study the absorption of clockwise polarized 
electromagnetic field by
the pair of electrons occupying the triangular ring. The corresponding absorption
coefficients are calculated in dipole and low temperature approximations according 
to the formula \cite{Chuang95}
\begin{equation*}
\label{absorption_coeff}
 \alpha(\hbar\omega) = {\cal{A}}\,\hbar\omega\sum_{f} 
 |\,\langle f|\bm{\varepsilon}\cdot\bm{d}|i\rangle\,|^2 \,
 \delta\!\left[\,\hbar\omega
-\left({\cal{E}}_{\mathrm{f}}-{\cal{E}}_{\mathrm{i}}\right)\,\right]\,,
\end{equation*}
where ${\cal{A}}$ is a constant amplitude, $\omega$ the frequency of the electromagnetic
field,
$\bm{\varepsilon}$ the unit polarization
vector, $\bm{d}$ is the electric dipole moment, 
and ${\cal{E}}_{\mathrm{i,f}}$ are the 
energies of the initial ($\vert i\rangle$) and final ($\vert f\rangle$)
many-body states.


In the numerical calculations we use InAs material parameters which are: 
$m_{\mathrm {eff}}=0.023$, $g_{\mathrm {eff}}=-14.9$, and 
the relative permittivity $\epsilon=15$.
The studied sample is an equilateral triangular 
ring of height and side thickness equal to $75$ and $10$ nm, respectively.

\vspace{12pt}
\noindent
\textbf{3. RESULTS}
\vspace{3pt}


\begin{figure}
\centering
 \includegraphics[width=1.0\textwidth,angle=0]{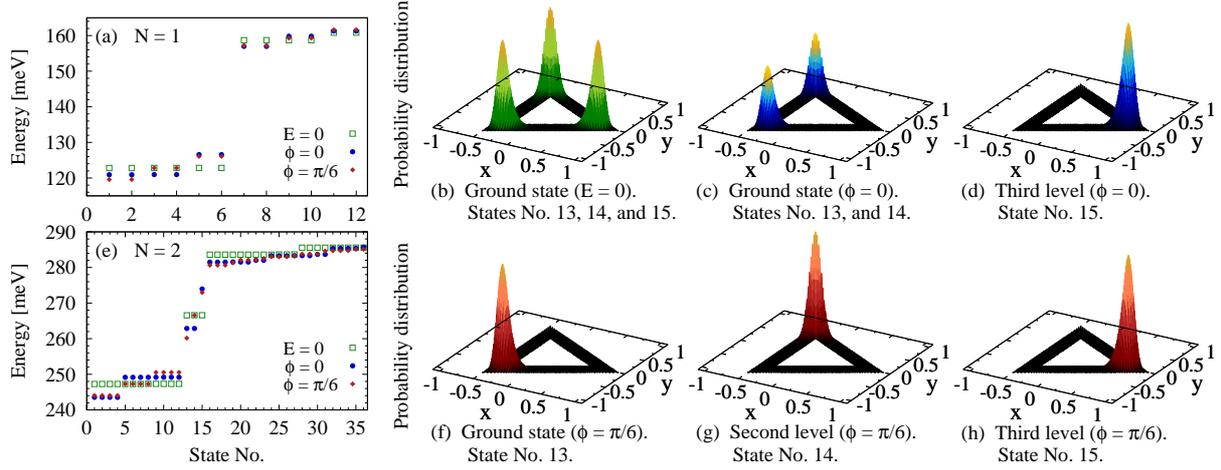}
\caption{\textit{Single-particle (a) and many-body (e) energy levels of 
         electrons confined in a triangular quantum ring in the absence 
         of external fields (green squares) and in the presence of an electric 
         field of amplitude $E=0.11$ mV/nm parallel to the $x$ axis 
         (blue circles) and one of the sides (red diamonds). The number of 
         particles ($N$) is shown in each figure. (b-d) and (f-h)
         probability distributions associated with some of the levels shown 
         in Figs. (a) and (e). The upper label refers to single-particle 
         quantities while the lower labels to many-body ones.}}
\label{FigEnergy}         
\end{figure}

\noindent
Polygonal quantum rings have unique physical properties which depend on the number
of corners and details of the sample symmetry. In particular, if there is a single 
electron confined in a symmetric (equilateral) sample, then its energy level degeneracy 
is determined solely by the number of vertices while the splittings between the 
levels depend on the aspect ratio between the side thickness and the radius of the 
external circumcircle. 
The ground state of an electron confined in a triangular ring is twofold 
(spin) degenerate and is followed by a sequence of alternating pairs of four- and 
twofold degenerate levels. The latter ones are degenerated due to spin, the 
fourfold degeneracy originates from spin and finite orbital momentum 
[Fig.\ \ref{FigEnergy}(a) green squares]. 
Low-energy electrons occupy areas between internal and external 
boundaries and for sufficiently thin rings are completely depleted from sides 
[Fig.\ \ref{FigEnergy}(b)]. These levels are energetically separated from higher 
(side-localized) states by a considerable energy gap which decreases with increasing 
the aspect ratio.
Contrary to square and hexagonal rings, for the triangular case this gap never diminishes to 
values comparable to splittings in the corner domain \cite{Sitek15,Sitek16}.

The degeneracy due to orbital momentum is lifted when the sample is exposed to an 
electric field parallel to the ring's plane. Irrespective of the field orientation, 
all the levels are spin degenerate. Beyond creating an  
energy splitting within each group of fourfold degenerate states, the field may also 
increase the splittings 
present in its absence. This stretching of the energy interval of every
set of six states is particularly strong for the corner-localized domain
[Fig.\ \ref{FigEnergy}(a)]. The field strongly rearranges
electron localization. If it acts along the $x$ axis then the two lowest levels
are quasi-degenerate and, due to the field direction, are shifted to lower energies
while the addition to the third level (states No. 5 and 6) is positive and slightly larger
[Fig.\ \ref{FigEnergy}(a) blue circles].
The two close by levels are associated with the same probability distributions which 
form two maxima at the ends of the side perpendicular to the electric field 
[Fig.\ \ref{FigEnergy}(c)]. Electrons excited to the third level occupy only the 
corner area with the highest value of the $x$ coordinate 
[Fig.\ \ref{FigEnergy}(d)]. Both probability distributions are symmetric with 
respect to the $x$ axis, i.e., orientation of the electric field.

If the external field is parallel to one of the sides then the six lowest states are 
arranged into three well separated levels [Fig.\ \ref{FigEnergy}(a) red diamonds]. 
In this domain the energy shifts are: positive and negative of the same magnitude for 
the corners located at the ends of the side parallel to the field, and negligibly small
for the opposite corner. The contribution to the second level does not vanish totally 
due to the finite areas between internal end external boundaries. The corresponding 
probability distributions form maxima around single corners which are determined by 
the direction and orientation of the external field with respect to 
the polygon [Figs.\ \ref{FigEnergy}(f)-\ \ref{FigEnergy}(h)].


Energy levels of two Coulomb interacting electrons confined in a triangular 
quantum ring are arranged in groups of quasi-degenerate levels. The fifteen lowest 
states are built of only corner-localized single-particle states, and thus are 
associated with probability distributions of this type [Fig.\ \ref{FigEnergy}(b)]. 
The ground state and 
the eleven consecutive states refer to particles distributed between two 
different corners. Due to the largest possible spatial separation, the 
contribution from Coulomb interaction to these states is the smallest. The 
opposite occurs for the smallest distance between particles, i.e., when 
the two electrons occupy the same corner area, which is possible for particles
in a spin singlet state. The three states of this kind (states No. 13, 14, and 15)
are shifted to much higher energies than other corner states. 
Since the spin of all three states is determined, their 
degeneracy is only of the orbital origin and resembles 
the arrangement of the lowest single-particle states with respect to spin. 
The higher states contain contributions from corner- and side-localized 
single-particle states, and thus are associated with mixed corner-side-localized
probability distributions. These states are separated from the lower group of corner 
states by an energy interval which is of the order of the gap separating corner- 
from side-localized single-particle states and is 'broken' only by the three in-gap
states [Fig.\ \ref{FigEnergy}(a) green squares] \cite{Sitek17}.


The external electric field induces large splittings within quasi-degenerate
groups of two-electron states. Like in the case of single-particle states, 
the most pronounced
effects occur for corner-localized states, in particular for the in-gap states.
If the field is parallel to the $x$ axis, then the twelve lowest states 
are split into two groups separated by $5.6$ meV, shifted to higher and lower
energies with respect to the states in the absence of the field. The energy interval
of in-gap states is also stretched. The two lowest states are nearly degenerate 
while the energy of the third one is $11$ meV higher
[Fig.\ \ref{FigEnergy}(a) blue circles]. As the energy levels,
the corresponding probability distributions also resemble single-particle ones.
This means that the two lowest in-gap states contain only contributions from
the single-particle ground state and the two states forming the second level,
while the third level is built of the fifth and sixth single-particle states 
[Fig.\ \ref{FigEnergy}(c) and\ \ref{FigEnergy}(d)].

An electric field parallel to one of the sides rearranges the twelve lowest 
states into three fourfold degenerate levels equally separated from each other 
by $3.3$ meV. In this case the energy of one of the in-gap states is decreased 
by $6.4$ meV while the energy of another one is increased by the same amount
with respect to the energies achieved in the absence of the field,
the energy of the third one is negligibly affected. 
A pair of electrons in one of these states occupies one particular corner, i.e.,
each of the in-gap states is built of only two spin degenerate single-particle 
states [Figs.\ \ref{FigEnergy}(f) -\ \ref{FigEnergy}(h)].


\begin{figure}
\centering
 \includegraphics[width=1.0\textwidth,angle=0]{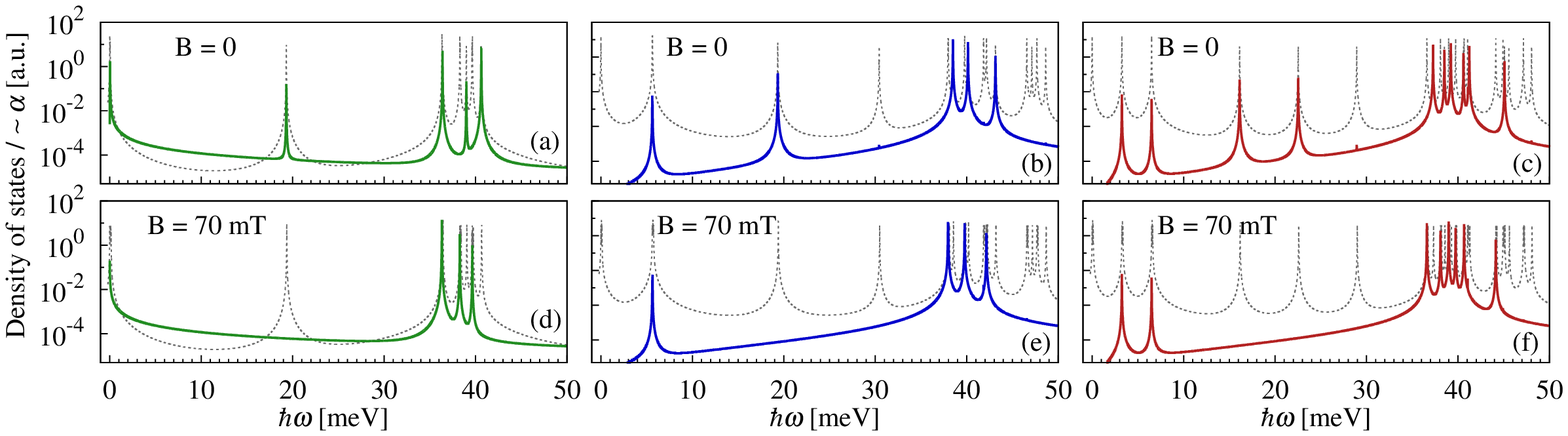}
\caption{\textit{Density of states (grey dotted) and 
         absorption coefficients associated with the excitation of 
         a ground state of a pair of Coulomb interacting electrons confined 
         in a triangular quantum ring: 
         in the absence of external fields (a), exposed to a static electric 
         field of amplitude $E=0.11$ mV/nm parallel to the $x$ axis (b), 
         and to one of the sides (c). (d-f) exposed to external electric fields 
         as in (a-c) and magnetic field of $70$ mT.}} 
\label{FigAbsorption}
\end{figure}

In this paper we do not take into account spin-orbit interaction, and thus the  
analysed optical transitions conserve spin. Circularly 
polarized electromagnetic field impinged on a sample containing a pair of 
Coulomb interacting electrons may excite the system to one of the nearby corner
states, to one of the in-gap states and to three states associated with mixed 
corner-side-localized probability distributions 
[Fig.\ \ref{FigAbsorption}(a)] \cite{Sitek17}. The number of optically induced 
transitions does not change when the sample is exposed to an external electric field 
parallel to the $x$ axis.
The ground state and the optically accessible 
in-gap state are associated with the same probability distributions 
[Fig.\ \ref{FigEnergy}(c)]. Due to the identical contributions from the electric 
field, the energy shifts of both these states are the same which 
results in conserved energy required to excite the system 
to the in-gap state. 
The field creates a splitting in the lowest group of corner states. This increases 
the separation between the ground state and the lowest optically excited state, 
and thus the excitation of this state requires higher energy than in 
the absence of the field [Fig.\ \ref{FigAbsorption}(b)]. The number of optical 
transitions increases when the external electric field is parallel to 
one of the sides. Such a field mixes the degenerate or quasi-degenerate optically 
active states (excited by a clockwise and counterclockwise polarized
electromagnetic field),
and rearranges them into pairs of well separated states which may be 
excited from the 
ground state irrespective of the polarization type.
In particular, it splits the lower in-gap states and breaks the previously 
forbidden gap with relatively distant states [Fig.\ \ref{FigAbsorption}(c)].

If the quantum ring is immersed in an external (static) magnetic field perpendicular to 
its plane, then all transitions to the in-gap states are blocked for relatively 
weak fields. All the in-gap states are well separated spin singlets which do not mix 
with other states for wide ranges of electric and magnetic fields. The opposite 
occurs within the group of twelve lower corner states. The ground state is either 
degenerate or separated from the second level by a very small energy interval. Due to the Zeeman 
splitting the degeneracy of all the levels is lifted. Moreover, for relatively 
weak fields the states originating from different levels may interchange. As a result 
the spin of the ground state changes, and thus the transition to the in-gap 
states is blocked [Figs.\ \ref{FigAbsorption}(d) -\ \ref{FigAbsorption}(f)].


\vspace{12pt}
\noindent
\textbf{4. CONCLUSIONS}
\vspace{3pt}

\noindent
We analysed energy levels and the corresponding electron localization of a single
electron and a pair of Coulomb interacting particles confined in a triangular 
quantum ring in the presence of an external electric field. We showed that the 
field lifts the degeneracies due to orbital momentum but conserves the twofold 
spin degeneracy of single-particle levels. Moreover, the field strongly 
redistributes particles within the ring. Depending on its orientation with respect 
to the cross section, the particles may be depleted from some regions of the sample. 

The energy spectrum of a pair of Coulomb interacting electrons is even more 
affected by the field. The degenerate and quasi-degenerate levels are split into 
well separated groups of states. In particular, the in-gap states, i.e., pairs of 
electrons in a spin singlet state occupying the same corner area, are stretched, 
and thus associated with energies belonging to a wide range of energies forbidden 
for non-interacting particles. Irrespective of the external electric field, 
the in-gap states degeneracy and the corresponding probability distributions 
reproduce those of single-particle corner states with respect to spin. 
The absorption of the many-body 
system depends on the orientation of the electric field. The excitation of the 
in-gap states is unaffected if the field is perpendicular to one of the sides, but 
the field parallel to one side allows for optical transitions to two 
distant in-gap states. 
The excitation of the highest two-particle corner-localized states is blocked in 
the presence of a weak magnetic field perpendicular to the ring's plane.

\vspace{12pt}
\noindent
\textbf{ACKNOWLEDGEMENTS}
\vspace{3pt}

\noindent
This work was financed by the Icelandic Research Fund.

\bibliographystyle{plain}

\end{document}